# Lithium doped graphene as spintronic devices

Narjes Kheirabadi [1,2]

[1] Department of Physics, Iran University of Science and Technology, Narmak, Tehran, Iran, 1684613114.

[2] Department of Physics, Lancaster University, Lancaster LA1 4YB, United Kingdom.

# Abstract

Generating spintronic devices has been a goal for the nano-science. We have used density function theory to determine magnetic phases of single layer and bilayer lithium doped graphene nanoflakes. We have introduced graphene flakes as single molecular magnets, spin on/off switches and spintronic memory devices. To aim this goal, adsorption energies, spin polarizations, electronic gaps, magnetic properties and robustness of spin-polarized states have been studied in the presence of dopants and second layers. We find that for bilayer SMMs with two layers of different sizes the highest occupied molecular orbital and the lowest unoccupied molecular orbital switch between the layers. Based on this switch of molecular orbitals in a bilayer graphene SMM, spin on/off switches and spintronic memory devices could be achievable.

# Introduction

Today, parallel to the developing of technology, molectronic, i.e. molecular electronic, is an avenue for search towards finding new applications of materials on a scale where quantum mechanics dominates electron kinetic behaviour [1, 2]. Meanwhile, single molecular magnets (SMMs), magnetic molecules with stable magnetization at room temperature, have a special role [3-7]. SMMs could be used as ferromagnetic materials (FM) which are the basis of spintronic devices, spin amplifiers, and those devices which magnetically store information at a molecular level. When a current passes through a SMM, the current will be spin polarized as do a spin amplifier. A very high spin polarized current with its magnetization parallel to the SMM magnetization flows for a time equivalent to the relaxation time (giant spin amplification) [8, 9]. For large currents, this process can lead to a selective drain of spins with one orientation from the source electrode, thus transfer a large amount of the magnetic moment from one lead to the other [4].

The high coherence time, the absence of conformational changes, weak spin-orbit and hyperfine interactions of carbon (C) atoms make the development of carbon based SMMs more desirable [4, 10]. Meanwhile, the hexagonal arrangement of carbon atoms in two dimensions, graphene based materials, have a special role. The low efficiency of the spin relaxation, for manipulated bilayer graphene up to a nanosecond [11-13], the scalability of the total spin, and its stability up to room temperature for single layer graphene render graphene as an excellent candidate for spintronic devices; such as spin memories, transistors, and qubits [14-17]. Furthermore, it has been proved that magnetic graphene nano-flakes (GNFs) hold the promise of an extremely long spin relaxation and decoherence times, with weak coupling between electron spins, and long-range magnetic order [18]. It is noteworthy that GNFs transport properties could be changed with the application of the electric [19] and magnetic fields [20], additional layers [21], and by controlling its geometry [22-25].

From the theoretical part of view, it is found that boron and nitrogen zigzag and armchair doped graphene nanoribbons could be FM [26]. Moreover, it is predicted that ZGNRs have a magnetic insulating ground state with FM ordering at each zigzag edge and antiparallel spin orientation between the two edges [27-29]. In addition, it has been predicted that the edge effect is of great importance for spin related properties of GNRs [30]. While, there are not many cases studied doped magnetic properties of GNF, especially for the case of bilayer ones. Consequently, the study of these SMMs is a missing part of molectronic [31]. Here, we have considered this part of theoretical study to plan a better magnet which makes better contacts with leads and wires in a circuit, or to give suggestions for isolating a material which works better at room temperature. Furthermore, control of unidirectional logic flow, preservation of the intrinsic properties for non-destructive readout of the spin states are all open issues [31]. We have determined a reasonable correlation between the quantum interference picture and orbital interaction pictures. The orbital view provides a better understanding of the intermolecular transport phenomena and connects the analysis of the wave function to the intuitive quantum interference effects [32], what we try to study in this article.

Furthermore, different methods have been developed to create graphene components with different shapes, sizes, and edge states [33-38]. Nano cutting, electro-beam lithography [39, 40] or $C_{60}$ transformation [41], heat-induced fractionalization of graphite [42], heat-induced conversion of nanodiamonds [43, 44] and silicon carbide [45], unzipping of carbon nanotube [46] are some methods to produce GNFs in laboratories. Furthermore, bilayer graphene could be defined by topgates [47], and it could be produced asymmetrically by use of an epitaxial growth method [48]. Consequently, the ability to modify the electronic properties of finite-size graphene by varying their size, shape, and edge orientation or defects is an important part of graphene based molectronic researches. While, a range of experiments from charge detection in bilayer GNFs and observation of spin states in GNF [20, 49] to observation of excited states in quantum dots [50, 51] all indicate that the spintronic devices based on the GNF are reachable by modern nano-scale fabrication methods.

## Systems Geometry and Computational Details

In this article, properties of single layer (Fig.1) and bilayer (Fig.3), Lithium (Li)-doped hexagonal shaped GNFs have been studied. Because of the predicted important role of alkali-metal decorated graphene [52] and specially Li [53, 54] we have used this element. The considered Li doped GNFs are also Hydrogen (H) terminated, in order to remove the effect of dangling bonds.

In the present work, we use whole-electron broken symmetry and first principles DFT calculations. The basis is 6-31g* and the hybrid exchange-correlation functional is B3LYP [55, 56] employing the Gaussian 03 software package [57] to verify the existence of the magnetic phases. The goal is answering some key questions about doped graphene magnetic properties. First, the robustness of the spin-polarized states will be studied in the presence of both impurities and of a second layer. The answer to this question is not only scientifically interesting for better understanding of the physical mechanism of spin polarization in hexagonal nanoflakes, but also it has important technological implications in the reliability of doped hexagonal nanoflake as a new class of SMM for spintronic materials [31]. Second, the author will discuss how the magnetic structures of hexagonal nanoflakes change with the size of layer, how it acts compare to infinitely hexagonal flakes, how change a doped Li-GNF to have a better magnetic ground state, and what are magnetic properties of bilayer hexagonal GNFs. Furthermore, band gaps, and magnetization of graphene flakes have been calculated as a function of second layer, and adsorbed Li distance from the centre of the flake. In addition, to study the graphene magnetization and its applications, molecular orbital theory will be employed. Finally, based on results of this study, some suggestions for the graphene based spintronic amplifier, the spin on/off switch, and the spin based memory device will be determined.

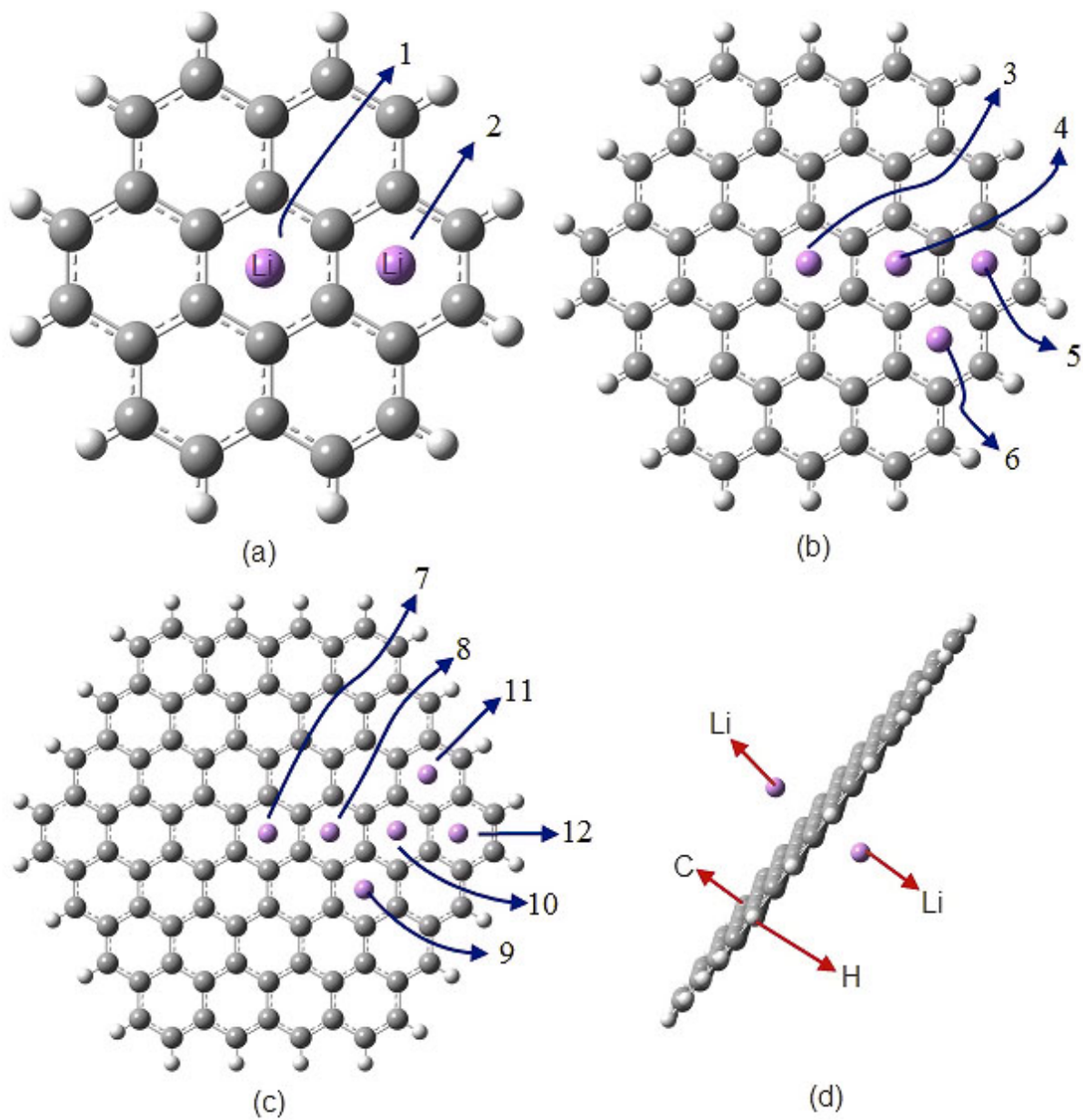

Fig. 1. Li Doped Single layer GNFs. C, H and Li atoms are respectively, grey, white and violent (a) Two different $LiC_{24}H_{12}$ flakes (number 1 and 2). (b) Four distinct $LiC_{54}H_{18}$ flakes (number 3, 4, 5 and 6). (c) Six dissimilar Li $C_{92}H_{24}$ flakes (flakes number 7 through 12). (d) $Li_2C_{92}H_{24}$, a flake number 13, doped by two Li atoms [58].

## Results

### Single layer GNFs

In this section, we have studied 13 doped GNFs (Fig.1). The flakes have been arranged by size and by the distance of Li atoms from the centre of the flake. The results of calculations are illustrated in the Fig. 2 and summarised in table.1.

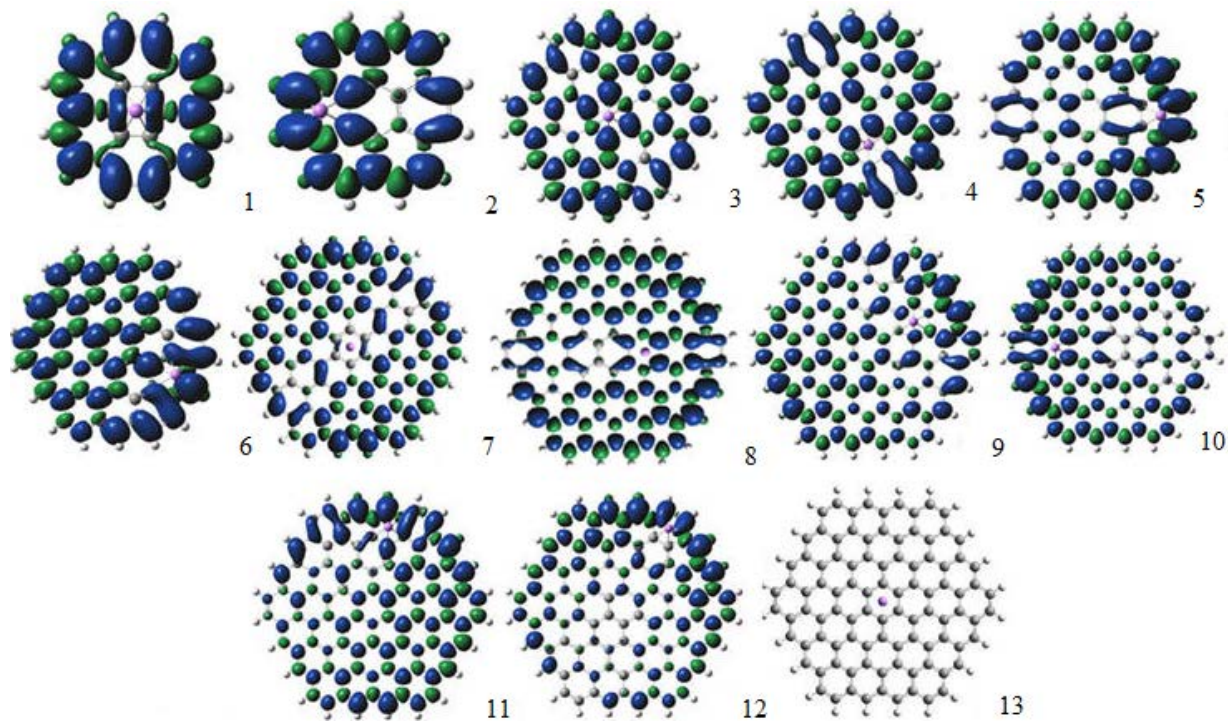

Fig.2. Studied single layer Li doped GNFs. The electron density from spin density is depicted in this figure; the isosurface value is ±0.0004 electrons per cubic atomic unit for positive values in the blue and negative values in the green.

The relative stability and the Highest Occupied Molecular Orbital-Lower Unoccupied Molecular Orbital (HOMO-LUMO) gap of some of these flakes have been studied before [59]. However, these properties for a larger group of Li doped graphene have been studied here. Flakes 5, 6, and 9 are newly considered flakes; the new results are consistent with results in Ref.[59].

According to Ref. [59], edge states are generally more stable. Here, this result has been confirmed for a larger group of Li doped graphene. To describe it in detail, according to the table.1, flakes NO.5 and NO.6, which have a Li atom nearer to edges, have adsorption energies which are, respectively, -0.024 and -0.0234 eV; more than flakes NO.3 (-0.017) and 4 (-0.017eV). In addition, for flake NO.9, the adsorption energy (-0.025eV) is slightly red shifted relative to flakes NO. 10 to NO.12 (-0.026, -0.032 and -0.032eV), while this flake has more adsorption energy relative to flake NO.7 and NO.8 (-0.023 and -0.024eV) whose adsorbent is nearer to the edge.

Concerning spin polarization, the single electron of a Li atom breaks the symmetry between spin up and spin down states. As is clear in Fig. 2, all doped graphenes with one Li atom are FM [15] while the last one, which is a flake doped by two Li atoms, is NM. The NM flake does not have any localized spin polarized states even at the edges. While the ultimate goal of the use of graphene in the next generation electronic is to realize all-graphene circuits with functional devices built from graphene layers or graphene flakes [30, 60], graphene doped Li has an additional advantage. It is possible to build a unique circuit of both FM and NM phases by use of Li-doped graphene. Because they have the same hexagonal carbon structure, excellent connections in a circuit are predictable.

For a better spin polarized current conduction, to join in a circuit, a contact group should be attached to localised orbitals [61]. As it is shown in Fig.1, in flakes with Li adsorbed on the edge, where spin polarization is localized, the middle part has less spin polarized regions, especially when the Li atom is adsorbed on a carbon ring which has two H atoms as the first neighbours. When, the Li atom is adsorbed on a symmetry line pass through those benzene rings with two Hydrogen atoms in both sides, around this line, distribution of spin up density is stronger than spin down density (so that spin up states in the centre make a tunnel-like spin up zone). As for the gap, all of the flakes doped by one Li atom are spin polarized and the alpha gap is red shifted by increasing of the flake size; as is the beta gap. To describe it in detail, according to Fig.1, the alpha gap for flakes NO.5, 6 and 9 are 1.02, 0.97, 0.68eV, respectively. While, the beta gaps are 2.67, 2.68 and 1.97eV, respectively. Consequently, our results confirm previous results about gap change according to flake size and the distance from the edge states [59].

Because of the different spin distribution population through surface of flakes, each flake has different FM properties. It is noteworthy that the difference between the highest spin polarization and the lowest spin polarization of C atoms has been selected as a factor to evaluate the flakes spin polarization (table 1).

Table 1: Monolayer Muliken spin density variation interval. The difference between the highest spin polarization and the lowest spin polarization of C atoms has been used to evaluate the spin polarization of flakes. Flake No.0 is a benzene ring doped by one Li atom in the centre.

| Flake NO. | 0 | 1 | 2 | 3 | 4 | 5 | 6 | 7 | 8 | 9 | 10 | 11 | 12 |
|---|---|---|---|---|---|---|---|---|---|---|---|---|---|
| Polarization (Mulliken atomic spin densities) | 0.12 | 0.28 | 0.27 | 0.25 | 0.26 | 0.30 | 0.30 | 0.20 | 0.33 | 0.26 | 0.24 | 0.25 | 0.33 |

According to table.1, flakes which have a Li atom nearer to boundary H atoms are better FM. For instance, the Li atom in flake NO.7 is adsorbed in the centre (Fig.2), far from edges; so; it has the least amount of the magnetization of all of flakes (table.1). These results show that increasing of the temperature which moves Li to the edge [62] is a factor which increase FM properties. As a result, FM properties of suggested amplifiers could act well at room temperature. In addition, the flake size is another important factor which affects magnetization. Larger flakes have generally a stronger distribution of spin through the flake [4]. By these considerations, larger flakes with Li at the edge, especially when Li is adsorbed on a ring which has two H atoms, are the best spin amplifiers. This note is a good reason to use a Li doped graphene as the spin amplifier; such structures are also more stable.  For example, between flakes NO. 2, 5 and 12 which have a Li atom at the edge, flake NO.2 whose C atoms are less is less spin polarized, while flake NO.12 has the largest size and the largest amount of spin polarization. Consequently, because of the graphenes high spin life time and Li-doped graphene FM properties, all the above flakes are predicted to be room temperature SMMs. Furthermore, only spin states parallel to the molecular magnetization can flow through these flakes. It is also predictable that the spin polarized current display a very high spin polarization for a time equivalent to the relaxation time of graphene. SMM NO.12 is the best candidate for gigantic spin amplification and SMM applications. Consequently, SMMs and similarly giant spin amplifications are achievable by use of GNFs. However, the possibility of having a better room temperature spin amplifier based on even larger graphene flakes is predictable.

## Bilayer Graphene Flakes

In the next step, to consider effects of an additional layer on magnetic properties of GNFs, a second layer has been added to single layer flakes. These bilayer structures have been arranged according to the number of C atoms and the Li distance from the centre (Fig.3). According to Fig.3, we have studied bilayer flakes with both layers of different and of equal size. Moreover, different positions for an adsorbent Li atom and the second layer have been considered. Results for these flakes have been gathered in Fig.3 and table2. It is worthy to mention that, because DFT is based on whole electron calculations, the radius of the largest-considered monolayer GNF is larger than that of the studied bilayer GNFs.

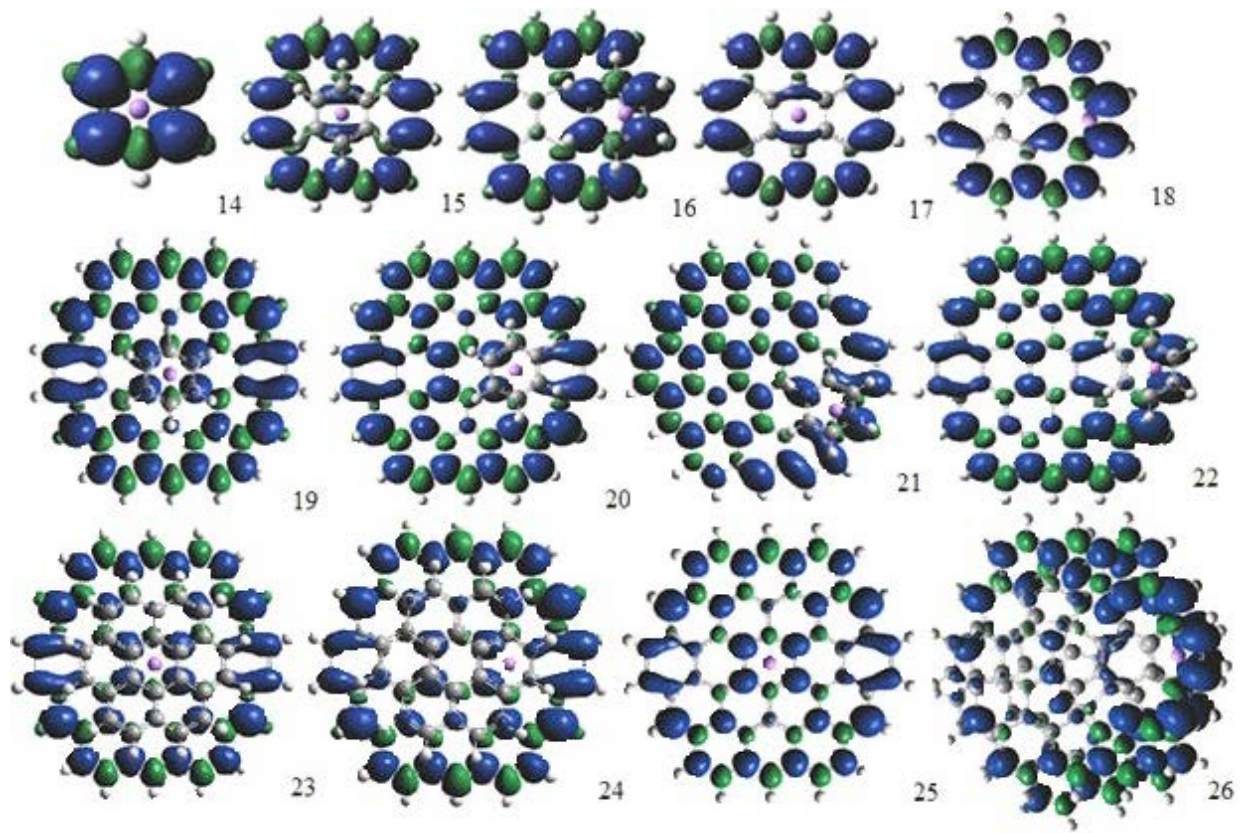

Fig.3. Studied bilayer GNFs. The isosurface value is ±0.0004 electrons per cubic atomic unit for positive values in the blue and negative values in the green.

Table 2. Results of DFT calculations for bilayer GNFs. These results for the adsorption energy, the spin polarization, and alpha and beta gaps of bilayer GNFs have been gathered in this table.

| NO. | Adsorption Energy (eV) | Spin-polarization (Mulliken atomic spin densities) | Alpha/Beta Gap (eV) |
|---|---|---|---|
| 14 | -0.027 | 0.22 | 1.32/5.86 |
| 15 | -0.034 | 0.28 | 0.73/3.28 |
| 16 | -0.039 | 0.23 | 1.1/3.67 |
| 17 | -0.033 | 0.14 | 0.82/3.54 |
| 18 | -0.043 | 0.13 | 0.87/3.57 |
| 19 | -0.050 | 0.20 | 0.79/2.62 |
| 20 | -0.049 | 0.24 | 0.82/2.62 |
| 21 | 1.444 | 0.29 | 0.91/2.67 |
| 22 | -0.053 | 0.29 | 0.95/2.65 |
| 23 | -0.043 | 0.20 | 0.64/2.49 |
| 24 | -0.047 | 0.25 | 0.61/2.47 |
| 25 | -0.046 | 0.10 | 0.61/2.51 |
| 26 | -0.046 | 0.15 | 0.60/2.59 |

The arrangement of flakes based on their stability is 21, 14, 17, 15, 16, 23, 18, 25, 26, 24, 20, 19 and 22. For relative stability, a particular rule hasn't been found. Unlike single layers, the total number of C atoms is not a determining factor for the amount of adsorption energy. For instance, flake NO.22 has less C atoms relative to Flake NO.24 and NO.25, but its adsorption energy is red shifted. In contrast, flake NO.18, which has more C atoms relative to flakes NO.16 and NO.15, is more stable than both of those. However, between similar flakes, those which are doped by a Li atom and a benzene ring nearer to the edge are more stable. According to table 2, between flakes 23 and 24, the later, whose Li atom is nearer to the edge is more stable; same about the flake NO.15 relative to 16 and NO.25 relative to 26. Flake NO.22 is more stable than flake NO.20, and flake NO.20 adsorption energy is red shifted relative to the flake No.19, similar to the flake NO.16 relative to the flake NO.15. It is noteworthy that flake NO.21 is not stable exceptionally. This has happened because of effects of interactions between boundary H atoms in two layers. In this case, H atoms in boundaries have adsorbed each other strongly. While, in another stable flake, NO.22, H atom bonds in the edge of one layer are parallel to those of another layer. In this flake the benzene ring and the Li atom are both in the edge state. Consequently, it is not possible to have a stable doped bilayer GNF similar to flake NO.21.

According to table.2, the arrangement of spin polarization amount according to the flake number is: 25, 18, 17, 26, 19, 23, 14, 16, 20, 24, 15, 22, and 21. According to this table, the spin polarization interval are generally smaller for bilayer GNFs relative to monolayer ones with a similar radius; for GNFs NO.16 to 26 spin polarization is less than a similar doped single layer GNF. Furthermore, according to Fig.3, in the flake NO.14 which has a Li atom between two coupled benzene rings, both surfaces are ferromagnetically spin polarized and strongly correlated (Fig.4). For the flake NO.15, one layer is FM while the smaller layer does not have any spin polarization with $10^{-2}$ accuracy. Consequently, we have both FM and NM layers in the flake No.15. For the flake NO.16, like No.14, two surfaces are strongly dependent, so that the two surfaces have even a common joint for spin density regions as it is depicted in Fig.4. This strong entanglement between two layers happens because of the high concentration of the spin density at the edge. Flake number 17 has two similar flakes doped by one Li in the centre. The important note about this structure is a drastic reduction of the spin density relative to the similar monolayer Flake. However, both layers are still FM. GNF NO.18 is similar to NO.17; a reduction of magnetization in FM layers is noticeable. In fact for all GNFs, both doped layers are FM, if it be spin polarized. According to table 2 and Fig.3, flake NO.22 is more spin polarized relative to flakes NO.19 to 21, and the same rule applies for the flake NO.24 relative to 23; and the flake NO.26 relative to NO.25. Consequently, the spin polarization is dependent on the second layer size and position. In detail, edge adsorption for the second layer increases FM properties. However, according to Fig. 4 and table.2 for small size flakes, NO.14, 16 and 18, when layers are both small, because of the increase of the spin distribution on the edge, two layers are entangled in each other and this rule is not correct any more.

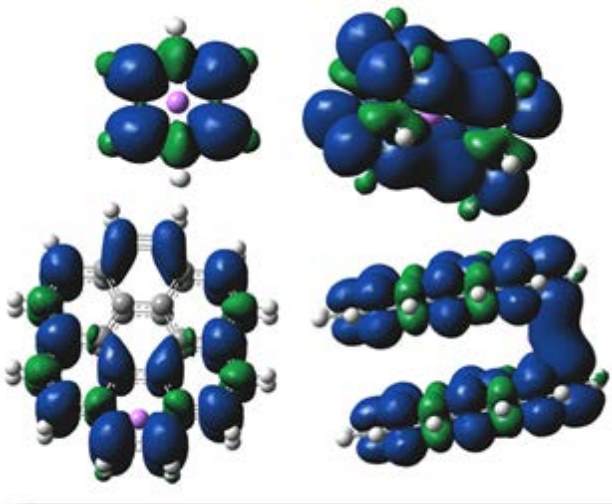

Fig 4. Different views for GNFs NO. 14 and 18. As it is visible in the right side of this picture (top and down), the spin density distribution for two different layers have been entangled similar to a joint. The isosurface value is ±0.0004 electrons per cubic atomic unit.

According to table 2, there are two effective factors influence bilayer GNFs gap amounts. First, the gap is generally red shifted by increasing the number of C atoms. For instance, between flakes NO.22 and 23, the gap of the second flake is red shifted, because it has more C atoms. The second factor which affects the gap is the second layer position. For instance, between flakes 19 to 22, the later which has benzene layer nearer to the edge has larger gap, same about flake NO.15 and 16. However, a systematic relation between the Li position and the gap has not been found.

In addition, according to Fig.3, similar to monolayer cases, if a Li atom is adsorbed in a ring which has two H atoms or Li is on a symmetry line passing through such a ring, a tunnel of spin polarized density is visible in the SMM. Consequently, those benzene rings with two H atoms at the edge are the best sections to make contacts, because those states have a strong concentration of spin up. In addition, a tunnel of spin polarized regions happens in that vicinity.

# Application

## Suggestions for spin amplifiers

As mentioned earlier, Li-doped single layers are generally better FM than bilayer graphenes and consequently those are better amplifiers. However, bilayer GNFs, the second layer could be managed for appropriate ferromagnetism and spin polarization. Between all bilayer, GNFs those which have a higher amount of adsorption energy and larger spin polarization are preferable for use as spintronic amplifiers. Consequently, the author suggests flakes 22 and 24 as spin amplifiers, because these flakes have both high spin density and high relative stability. The spin polarization of these bilayer GNFs is near to similar single layer GNFs.

The effect of temperature on adsorbed Li, at room temperature (270 < T < 400 K), is that Li atoms fall in the boundary condition [62]. For bilayer cases, the effect of increasing of temperature to control suggested spin amplifier is similar to the single layer. Moreover, on one hand, this research shows that those structures with benzene or Li at the edge are more stable relative to those structures that have benzene and Li in the centre. On the other hand, based on previous research [59], between two layers and in the middle of benzene rings are preferable for adsorbent Li atom. Consequently, by increasing of the temperature, the dopant moves to edge states of GNFs, where these GNFs are stable up to room temperature. Consequently, Li doped GNFs are SMMs and room temperature spin amplifiers. In addition, if we consider bilayer GNF as an amplifier, this type of amplifier will act even better at higher temperature. In fact, because of the migration of Li and benzene ring to the edge, the spin polarization and the stability of GNFs increase generally. However, the study of managing edge states, of adsorbent type to increase spin density and whether more odd numbers of Li atoms increases spin polarization needs more research.

## Spintronic on/off switch based on Li doped bilayer

Molecular orbitals (MOs) are conduction channels for electrons. These channels could be obstructed (localized) or not (delocalized); and simultaneously, those could be occupied by electrons or not. Any factor which changes this occupation allows us to tailor the electrical behaviour of the molecule [61]. MOs for GNFs are conducting channels; conversely, a non-conducting channel is a localized MO, which cannot connect both ends of the molecule attached to metallic contacts. Furthermore, shapes of frontier molecular orbitals explain qualitatively the conduction of electrons through molecules attached to macroscopic contacts [61]. Based on this analysis, the author suggests a structure similar to Fig.5 for applications in the field of a single electron on/off switch. According to Fig.5, bilayer GNFs which have layers of different sizes and a Li atom at the centre or an edge could act as a single electron on/off switch. In such a SMM, HOMO concentration is on the larger flake, while the LUMO concentration is on the smaller one. Such a switch of molecular orbitals between layers happens because of the repulsive effect of H atoms. It is worthy to mention that this structure stability and spin polarization in comparison with all other GNFs is high as discussed before. In addition, to have a single electron on/off switch for spintronic purposes, contacts should be added to the larger surface which has stronger magnetic properties and it is a stronger spin amplifier. Furthermore, based on table 2 and Fig.5, it is predictable that this type of switch will have a high on/off ratio because of the low covering effects of two doped layers with different size. However, this ratio will be decreased by the movement of Li atom to the edge of a SMM at higher temperatures (Fig.3).

**Memory devices based on Li doped bilayer**
According to Fig.5, when an electron transfers to the LUMO state of the depicted SMMs, there is no MO on the larger surface. This could be defined a "0" state for memory. On the other hand, when there are no electrons in the HOMO level, the probability of existence of electron in smaller surface is zero. This effect could be defined as a "1" digit for a relative memory device. According to Fig.5, for HOMO-2, HOMO-1, HOMO, LUMO, LUMO+1 and LUMO+2 such a switch of states is the same as for HOMO-LUMO states. Consequently, it is predictable that in such a bilayer SMM one layer will be responsible for conduction. Consequently, conduction channels switch between two layers and this up and down states could be used to define a memory device. Consequently, the author suggests the usage of bilayer graphene SMM as a memory device, as well.

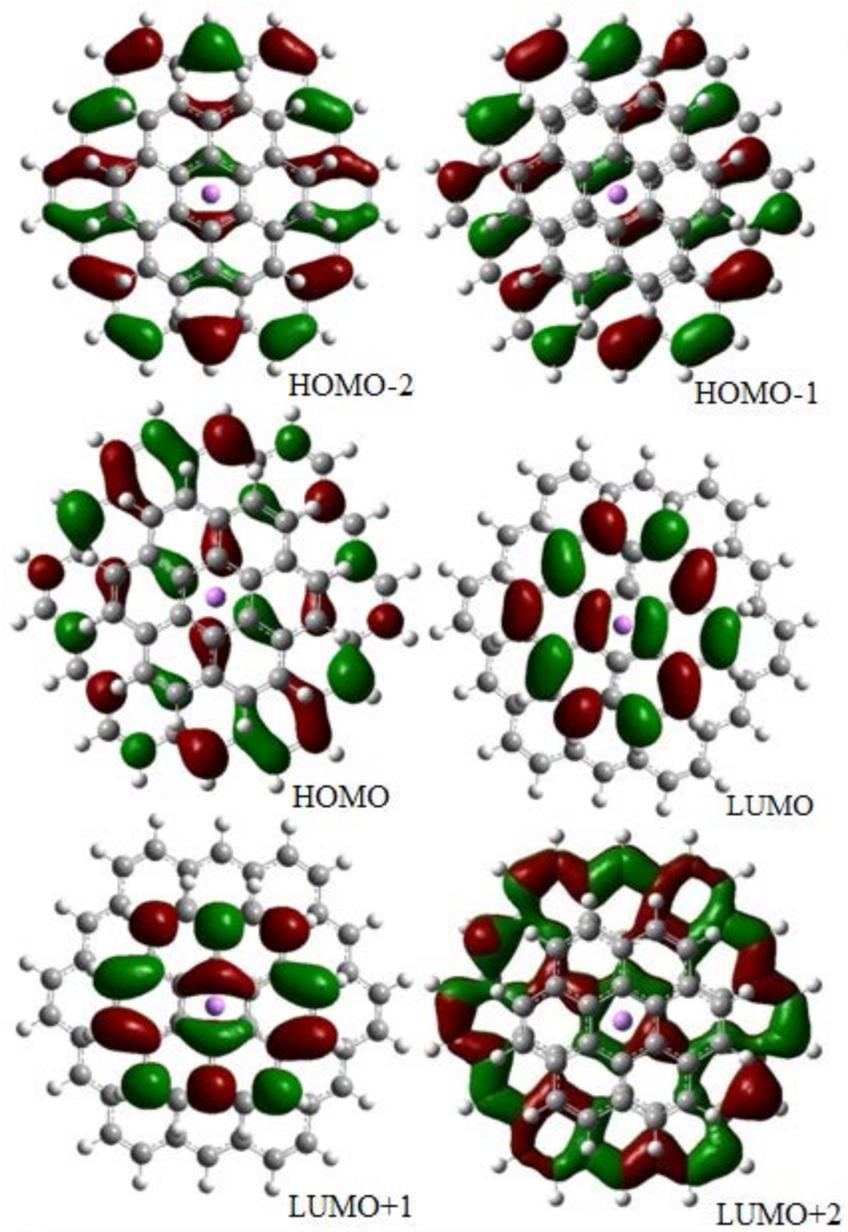

Fig.5. Spin up molecular Orbitals view for bilayer GNF NO.17 which is a doped bilayer graphene with layers of different sizes. According to this figure, for MO between HOMO-2 to LUMO+2 MOs are distributed on one layer. Colour code: red, positive; green, negative. The Isovalue is 0.02.

## Conclusion

All of the considered doped graphene structures with one Li atom are FM, while a flake doped by two atoms is NM and does not have any localized spin polarized state. This point makes it possible to build graphene based FM and NM by use of GNFs. Li doped GNFs are spin polarized and alpha and beta gaps are red shifted by increasing of the size. Furthermore, Li doped GNFs could be used as spintronic SMMs for room temperature applications, because these flakes generally have more stable edge states. For better conduction of spin polarized current, a contact group in a circuit should be attached where two H atoms are bonded in a benzene ring. In addition, larger SMMs which have a dopant nearer to the boundary are better FM. For, those flakes in which the dopant is adsorbed on a ring with two hydrogen atoms, spin amplifiers could act better, and this is a positive reason to use the doped GNFs as SMMs and spin amplifiers at room temperature.

For bilayer GNFs, flakes of different sizes have been studied, where the dopant position and the second layer position and size both are variable. Concerning relative stability, for similar flakes those flakes whose adsorbent is nearer to the edge are more stable. In addition, spin polarization is dependent on the second layer size and position. Edge adsorption of second layer increases FM properties, while layers are FM, if it be spin polarized. Furthermore, where two layers are small, because of the high intensity of the spin distribution on the edge, two layers are entangled in each other and spin polarization decreases drastically. Additionally, for bilayer SMMs, the alpha and beta gaps are generally red shifted by the increase of the number of C atoms. Moreover, those benzene rings with two H atoms at the edge are the best choices to be used as the contact group in a circuit.

For applications, single layer doped Li SMMs are better FM and also amplifiers. Because of high spin density and relative stability of flakes NO.22 and 24, SMMs similar to those flakes are predictable to act better as spin amplifiers. In addition, graphene SMMs which have different size could act as an on/off switch and better amplifiers. In such flakes, spin polarized MO switches between layers. Accordingly, the switch of molecular orbitals between layers could be defined as "0" and "1" states for molecular memory devices. Based on this point, the usage of bilayer SMMs as a memory device has been suggested.

## Acknowledgements

This work was made possible by the facilities of Computational Nanotechnology Supercomputing Centre, Institute for Research in Fundamental Science (IPM). The author is immensely grateful to Dr. Edward McCann for his precious comments on an earlier version of the manuscript.